\documentclass[%
 reprint,
 amsmath,amssymb,
 aps,
prb,
]{revtex4-1}

\usepackage{xcolor}
\usepackage{graphicx}
\usepackage{dcolumn}
\usepackage{bm}
\usepackage{mathrsfs}
\usepackage{hyperref}
\usepackage[normalem]{ulem}
\usepackage{lipsum}
\usepackage{IEEEtrantools}
\usepackage[version=3]{mhchem}
\usepackage{verbatim}
\usepackage{siunitx}
\usepackage{enumitem}
\usepackage{textcomp}
\bibpunct{[}{]}{;}{n}{}{}

\setlist[itemize]{noitemsep, topsep=0pt}
\begin{document}

\title{The NOMAD Artificial-Intelligence Toolkit:\\ Turning materials-science data into knowledge and understanding}


\author{Luigi Sbailò$^{1,2*}$, Ádám Fekete$^1$, Luca M. Ghiringhelli$^{1,2*}$, and Matthias Scheffler$^2$}
\affiliation{
$^1$Physics Department and IRIS Adlershof of the Humboldt-Universität zu Berlin, Germany \\
$^2$The NOMAD Laboratory at the Fritz Haber Institute of the Max-Planck-Gesellschaft and IRIS Adlershof of the Humboldt-Universität zu Berlin, Germany; \\ 
$^*$email: ghiringhelli@fhi-berlin.mpg.de; sbailo@fhi-berlin.mpg.de; 
}
\date{\today}

\begin{abstract}
We present the Novel-Materials-Discovery (NOMAD) Artificial-Intelligence (AI) Toolkit, a web-browser-based infrastructure for the interactive AI-based analysis of  materials-science findable, accessible, interoperable, and reusable (FAIR) data. The AI Toolkit readily operates on the FAIR data  stored in the central server of the NOMAD Archive, the largest database of materials-science data worldwide, as well as locally stored, users' owned data. The NOMAD Oasis, a local, stand alone server can be also used to run the AI Toolkit. By using Jupyter notebooks that run in a web-browser, the NOMAD data can be queried and accessed; data mining, machine learning, and other AI techniques can be then applied to analyse them. This infrastructure brings the concept of reproducibility in materials science to the next level, by allowing researchers to share not only the data contributing to their scientific publications, but also all the developed methods and analytics tools. Besides reproducing published results, users of the NOMAD AI toolkit can modify the Jupyter notebooks towards their own research work.

\end{abstract}
`
 \maketitle

\section{Introduction}

Data-centric science has been identified as the 4th paradigm of scientific research.
We observe that the novelty introduced by this paradigm is two-fold. First, the creation of large, interconnected databases of scientific data, which are more and more expected to comply with the so-called FAIR principles \cite{wilkinson2016fair} of scientific data management and stewardship: i.e., data and related metadata need to be findable, accessible, interoperable, and reusable (or repurposable, or recyclable).
The second aspect is the massive use of artificial-intelligence (AI) algorithms, applied to scientific data, in order to find patterns and trends that would be hard if possible at all to identify by unassisted human observation and intuition. 

Materials science has taken up in the last few years both aspects. Databases, in particular from computational materials science, have been created via high-throughput screening initiatives, mainly boosted by the US Materials-Genome Initiative, starting in the early 2010's, e.g., AFLOW \cite{curtarolo2012aflowlib}, the Materials Project \cite{jain2013commentary}, and OQMD \cite{saal2013materials}. 
At the end of 2014, the NOMAD (Novel Materials Discovery) Laboratory has launched the NOMAD Repository \& Archive
\cite{draxl2018nomad,draxl2019nomad,draxl2020big}, the first FAIR storage infrastructure for computational materials-science data. 
NOMAD's servers and storage are hosted by the Max Planck Computing and Data Facility (MPCDF) in Garching (Germany). The NOMAD Repository stores, as of today, input and output files from more than 50 different atomistic ({\em ab initio} and molecular mechanics) codes. It totals more than 100 million 
total-energy calculations, uploaded by various materials scientists from their local storage or from other public databases.
The NOMAD Archive stores the same information, but converted, normalized, and characterized by means of a metadata schema, the {\em NOMAD Metainfo} \cite{ghiringhelli2017towards}, which allows for the labeling of most of the data in a code-independent representation. The translation from the content of raw input and output files into the code-independent {\em NOMAD Metainfo} format makes the data ready for AI analysis. \\
Besides the above mentioned databases, other platforms for the open-access storage and access of materials science data appeared in recent years, such as the Materials Data Facility \cite{blaiszik2016materials,blaiszik2019data} and Materials Cloud \cite{talirz2020materials}. Furthermore, many groups have been storing their materials science data on Zenodo (\cite{zenodo_cite}), and provided the digital object identifier (DOI) to openly access them in publications. The peculiarity of the NOMAD Repository \& Archive is in the fact that users upload the full input and output files from their calculations into the Repository and then such information is mapped onto the Archive, which (other) users can access via a unified API.

Materials science has embraced also the second aspect of the 4th paradigm, i.e., AI-driven analysis. The applications of AI to materials science span two main classes of methods. One is the modeling of potential-energy surfaces (PES) by means of statistical models that promise to yield {\em ab initio} accuracy at a fraction of the evaluation time \cite{lorenz2004representing,behler2007generalized,bartok2010gaussian,bartok2013representing,schutt2017quantum,xie2018crystal} (if the CPU time necessary to produce the training data set is not considered). The other class is the so-called materials informatics, i.e., the statistical modeling of materials aimed at predicting their physical, often technologically relevant properties \cite{rajan2005materials,pilania2013accelerating,ghiringhelli2015big,isayev2015materials,ouyang2018sisso,jha2018elemnet}, by knowing limited input information about them, often just their stoichiometry. 
The latter aims at identifying the minimal set of descriptors (the materials' genes) that correlate with properties of interest. This aspect, together with the observation that only a very small amount of the almost infinite number of possible materials is known today, may lead to the identification of undiscovered materials 
that have properties (conductivity, plasticity, elasticity, etc.) superior to the known ones.

The NOMAD CoE has recognized the importance of enabling the AI analysis of the stored FAIR data and has launched the NOMAD AI Toolkit. This web-based infrastructure allows users to run in a web-browser computational notebooks (i.e., interactive documents that freely mix code, results, graphics, and text, supported by a suitable virtual environment) for performing complex queries and AI-based exploratory analysis and predictive modeling on the data contained in the NOMAD Archive.
In this respect, the AI Toolkit pushes to the next, necessary step the concept of FAIR data, by recognizing that the most promising purpose of the FAIR principles is enabling AI analysis of the stored data. As a mnemonic, the next step in FAIR data starts by upgrading its meaning to: Findable and {\em AI-Ready} data \cite{Scheffler2022}.

The mission of the NOMAD AI Toolkit is three-fold:
\begin{itemize}[leftmargin=10pt]
    \item Providing an API and libraries for accessing and analysing the NOMAD Archive data via state-of-the-art (and beyond) AI tools.
    \item Providing a set of shallow-learning-curve tutorials from the hands-on introduction to the mastering of AI techniques.
    \item Maintaining a community-driven growing collection of computational notebooks, each dedicated to an AI-based materials-science publication. By providing both the annotated data and the scripts for their analysis, students and scholars worldwide are enable to retrace all the steps that the original researchers followed to reach publication-level results. Furthermore, the users can modify the existing notebooks and quickly checks alternative ideas.\\
\end{itemize}
The data science community has introduced several platforms for performing AI-based analysis of scientific data,  typically by providing rich libraries for machine-learning and artificial intelligence and often offering users online resources for running notebooks. General-purpose frameworks such as Binder \cite{ragan2018binder} and Google Colab \cite{googlecolab}, as well as materials-science dedicated frameworks such as nanoHUB \cite{klimeck2008nanohub}, pyIron \cite{pyiron-paper}, AiidaLab \cite{yakutovich2021aiidalab}, and MatBench \cite{dunn2020benchmarking} are the most used by the community. In all these cases, a big effort is devoted to education via online and in-person tutorials. The main specificity of the NOMAD AI toolkit is in connecting within the same infrastructure the data, as stored in the NOMAD Archive, to their AI analysis. Moreover, as detailed below, users have in the same environment all available AI tools as well as access to the NOMAD data, without need to install anything.

This paper is structured as follows. In section II, we describe the technology of the AI Toolkit. In sections III and IV, we describe two exemplary notebooks. One notebook is a tutorial introduction to the interactive querying and exploratory analysis of the NOMAD Archive data. The other notebook demonstrates the possibility to report publication-level materials science results \cite{cao2020artificial}, while enabling the users to put their hands on the workflow, by modifying the input parameters and observing the impact of their interventions. 


\section{Results}

{\bf Technology} 

\begin{figure}
    \centering
    \includegraphics[width=\columnwidth]{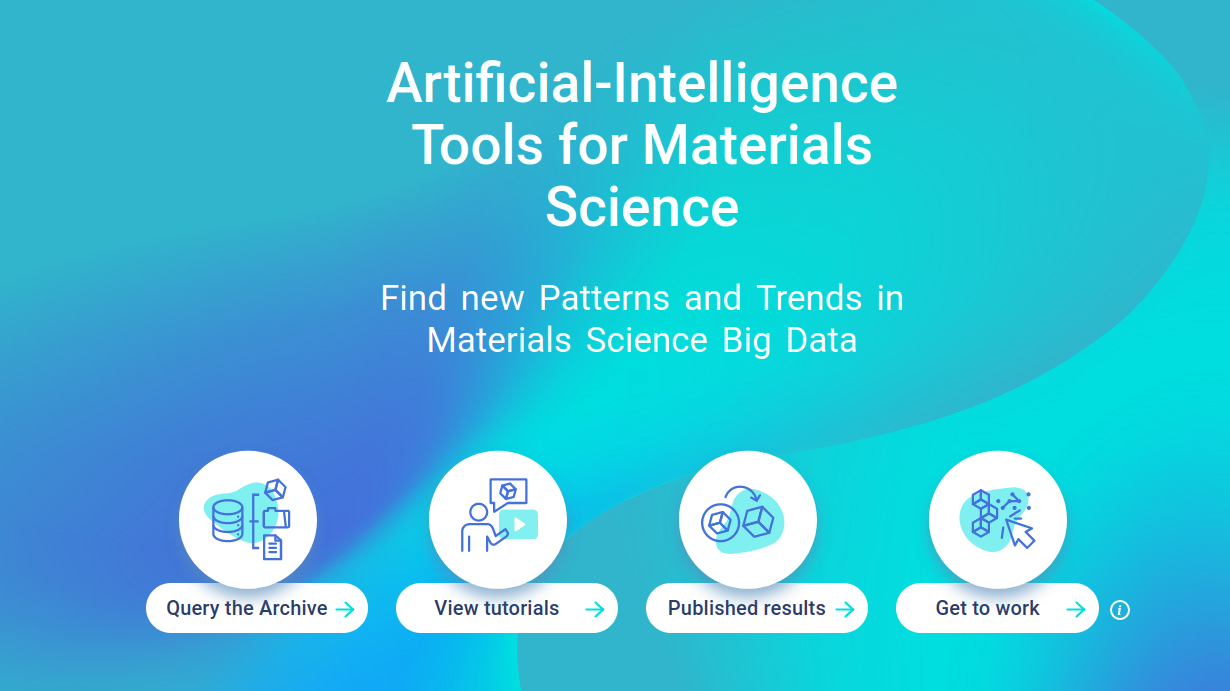}
    \caption{Home page of the NOMAD Artificial-Intelligence Toolkit, showcasing its three purposes: Querying (and analyzing) the content of the NOMAD Archive, providing tutorials for AI tools, and accessing the AI workflow of published work. The fourth access point, get to work, is for experienced users, who can create and manage their own workspace. 
    }
    \label{fig:analytics_toolkit}
\end{figure}

We provide a user-friendly infrastructure to apply the latest AI developments and the most popular machine-learning methods to materials-science data. 
The NOMAD AI Toolkit aims to facilitate the deployment of sophisticated AI algorithms by means of an intuitive interface that is accessible from a webpage.
In this way, AI-powered methodologies are transferred to materials science. In fact, the most recent advances in AI are usually available as software stored on web repositories. However, these need to be installed in a local environment which requires specific bindings and environment variables.
Such an installation can be a tedious process, which limits the diffusion of these computational methods, and also brings in the problem of reproducibility of published results.
The NOMAD AI Toolkit offers a solution to this, by providing the software, that we install and maintain, in an environment that is accessible directly from the web.

Docker\cite{dockerURL} allows to install software in a container that is isolated from the host machine where it is running.
In the NOMAD AI Toolkit we maintain such a container, installing therein software that has been used to produce recently published results and taking care of the versioning of all required packages.
Jupyter notebooks are then used inside the container to interact with the underlying computational engine.
Interactions include the execution of code, displaying the results of computations, and writing comments or explanations by using markup language.
We opted for Jupyter notebooks because such interactivity is ideal for combining computation and analysis of the results  in a single framework.
The kernel of the notebooks, i.e. the computational engine that runs the code, is set to read Python.
Python has built-in support for scientific computing as the SciPy ecosystem and it is highly extensible, because it allows to wrap codes written in compiled languages such as \textsc{C} or \textsc{C++}.
This technological infrastructure is built using JupyterHub\cite{jupyterHubURL} and deploys servers that are orchestrated by Kubernetes on computing facilities offered by the MPCDF in Garching, Germany. Users of the AI Toolkit can currently run their analyses on up to 8 CPU cores, with up to 10 GB RAM. 

A key feature of the NOMAD AI Toolkit is that we allow users to create, modify and store computational notebooks where original AI workflows are developed. 
From the `Get to work' button accessible at \url{https://nomad-lab.eu/aitoolkit}, registered users are redirected to a personal space, where we provide 10 GB of cloud storage and where work can also be saved.
Jupyter notebooks, which are created inside the `work' directory in the users' personal space, are stored on our servers and can be accessed and edited over time.
These notebooks are placed in the NOMAD AI Toolkit environment, which means that all software and methods demonstrated in other tutorials can be deployed therein.
The versatility of Jupyter notebooks in fact facilitates an interactive and instantaneous combination of different methods.
This is useful if one aims to, e.g., combine different methods available in the NOMAD AI Toolkit in an original manner, or to deploy a specific algorithm to a dataset that is retrieved from the NOMAD Archive.
The original notebook, which is developed in the `work' directory, might then lead to a publication and the notebook be added to the `Published results' section of the AI Toolkit.

{\bf Contributing}

The NOMAD AI Toolkit aims to promote reproducibility of published results.
Researchers working in the field of AI applied to materials science are invited to share their software and install it in the NOMAD AI Toolkit.
The shared software can be used in citeable Jupyter notebooks, which are accessible online, to reproduce results that have been recently published in scientific journals. 
Sharing software and methods in a user-friendly infrastructure such as the NOMAD AI Toolkit can also promote the visibility of research and boost interdisciplinary collaborations.

All Jupyter notebooks currently available in the NOMAD AI Toolkit are located in the same Docker container, thus allowing transferability of methods and pipelines between different notebooks. 
This also implies that software employed is constrained to be installed using the same package versions for each notebook. 
However, to facilitate a faster and more robust integration of external contributions to the NOMAD AI Toolkit, we allow the creation of separated Docker containers which can have their own versioning. 
Having a separate Docker container for a notebook allows to minimize maintenance of the notebook, and it avoids further updates when e.g. package versions are updated in the main Docker container.

Contributing to the NOMAD AI Toolkit is straightforward, and consists of the following steps:
\begin{itemize}[leftmargin=10pt]
\item Data must be uploaded to the NOMAD Archive and Repository. Either in the public server (\url{https://nomad-lab.eu/prod/rae/gui/uploads}) or in the local, self-contained variant (see Sec. \ref{sec:localapp}).
\item Software needs to be installed in the base image of the NOMAD AI Toolkit.
\item The whole workflow of a (published) project, from importing the data to generating results, has to be placed in a Jupyter notebook. The package(s) and notebook are then uploaded to GitLab in a public repository (\url{https://gitlab.mpcdf.mpg.de/nomad-lab/analytics}), where the back-end code is stored. 
\item A DOI is generated for the notebook, which is versioned in GitLab. In the spirit of, e.g., Cornell University's arXiv.org, the latest version of the notebook is linked to the DOI, but all previous versions are maintained.
\end{itemize}
Researchers interested in contributing to the NOMAD AI Toolkit are invited to contact us for further details.

{\bf Data-management policy}

For maintenance reasons, NOMAD keeps anonymous-access logs for API calls for a limited amount of time. However, those logs are not associated with NOMAD users; in fact, users do not need to provide authentication to use the NOMAD APIs. We also would like to note that query commands used for extracting the data that are analyzed in a given notebook are part of the notebook itself, hence stored. This guarantees reproducibility of the AI analysis as the same query commands will always yield the same outcome, e.g., the same data points for the AI analysis.
Publicly shared notebooks on the AI-toolkit platform are required to adopt the Apache License Version 2.
Finally, we note that the overall NOMAD infrastructure, including the AI Toolkit, will be maintained for at least 10 years after the last data upload.

{\bf AI Toolkit App}

\label{sec:localapp}
In addition to the web-based toolkit, we also maintain an App that allows to deploy the NOMAD AI Toolkit environment\cite{aitoolkit-app} on a local machine.
This App employs the same graphical user interface (GUI) as the online version, in particular, the user accesses it via a normal web browser. 
However, the browser does not need to have access to the web and can therefore run behind firewalls.
Software and methods installed in the NOMAD AI Toolkit will deploy the users' personal computational resources. 
This can be useful when calculations are particularly demanding, and also when AI methods are applied to private data that should not access the web.
Through the local App, both the data on the NOMAD server as well locally stored data can be accessed. The latter access is supported by the NOMAD OASIS, the stand alone version of the NOMAD infrastructure\cite{OASIS}. 

{\bf Querying the NOMAD Archive and performing AI modeling on retrieved data}

The NOMAD AI Toolkit features the tutorial `Querying the archive and performing Artificial Intelligence modeling' notebook \cite{query-archive-notebook} (also accessible from the `Query the archive' button at \url{https://nomad-lab.eu/aitoolkit}), which demonstrates all steps required to perform AI analysis on data stored in the NOMAD Archive. These steps are the following: $(i)$ querying the data by using the RESTful API (see below) that is built on the {\em NOMAD Metainfo}; $(ii)$ loading the needed AI packages, including the library of features that are used to fingerprint the data points (materials) in the AI analysis; $(iii)$ performing the AI
training and visualizing the results. 

The NOMAD Laboratory has developed the NOMAD Python package, which includes a client module to query the Archive using the NOMAD API.
All functionalities of the NOMAD Repository and Archive are offered through a RESTful API, i.e. an API that uses HTTP methods to access data. 
In other words, each item in the Archive (typically a JSON data file) is reachable via a URL accessible from any web browser.\\

In the example notebook \cite{query-archive-notebook}, we use the NOMAD Python client library to retrieve ternary elements containing oxygen. 
We also request that the {\em ab initio} calculations were carried out with the VASP code, using exchange-correlation (xc) functionals from the generalized-gradient-approximation (GGA) family. 
In addition, to ensure that calculations have converged, we also set that the energy difference during geometry optimization has converged. 
As of April 2022, this query retrieves almost 8\,000 entries, which are the results of simulations carried out at different laboratories. 
We emphasize that in this notebook we show how data with heterogeneous origin can be used consistently for machine-learning analyses.

Here, we target the atomic density, that is obtained by a geometrically converged DFT calculation. 
The client module in the NOMAD Python package establishes a client-server connection in a so-called {\em lazy} manner, i.e. data are not fetched altogether, but with an iterative query. 
Entries are then iteratively retrieved, and each entry allows to access data and metadata relative to the simulation results that have been uploaded.
In this example, the queried materials are composed of three different elements, where one of the elements is required to be oxygen. 
From each entry of the query, we retrieve the converged value of the atomic density and the name and stoichiometric ratio of the other two chemical elements.
During the query, we use the {\em atomic features library} (see below) to add other atomic features to the dataframe that is built with the retrieved data.
Before discussing the actual analysis performed in the notebook, let us briefly comment on the {\em NOMAD Metainfo} and the libraries of input (atomic) features.

{\bf The NOMAD Metainfo}

The NOMAD API access to the data in the NOMAD Archive, which are organized by means of the {\em NOMAD Metainfo}, which is presented in Ref. \cite{ghiringhelli2017towards} and \cite{ghiringhelli2022community}. Here, we mention that it is a hierarchical and modular schema, where each piece of information contained in an input/output file of an atomistic simulation code has its own metadata entry. The metadata are organized in {\em sections} (akin to tables in a relational database) such as {\em System}, containing information on the geometry and composition of the simulated system, and {\em Method}, containing information on the physical model (e.g., type of xc functional, type of relativistic treatment, and basis set). Crucially, each item in any section (a column in the relational database analogy, where each data object is a row) has a unique name. Such name (e.g. `atoms', which is a list of the atomic symbols of all chemical species present in a simulation cell)
is associated with values that can be searched via the API. In practice, one can search all compounds containing oxygen by specifying \texttt{query=\{'atoms': ['O']\}} as argument of the \texttt{query\_archive()} function, which is the backbone of the NOMAD API.

{\bf Libraries of input features}

Together with the materials data, the other important piece of information for an AI analysis is the representation of each data point. A possible choice, useful for exploratory analysis, but also the training of predictive models,  is to represent the atoms in the simulation cell by means of their periodic-table properties (also called {\em atomic features}), e.g., atomic number, row and column in the periodic table, ionic or covalent radii, electronegativity. In order to facilitate access to these features, we maintain the \texttt{atomic\_collections} library, containing features for all atoms in the periodic table (up to $Z=100$), calculated via DFT with a selection of xc functionals. Furthermore, we have also installed the \textsc{matminer} package\cite{ward2018matminer}, a recently introduced rich library of atomic properties from calculations and experiment. In this way, all atomic properties defined in the various sources are available within the toolkit environment.

{\bf Example of exploratory analysis: Clustering}

We now proceed with the discussion of the showcase notebook, which performs an unsupervised-learning analysis called clustering.
The evolutionary human ability to recognize patterns in empirical data has led to the most disparate scientific findings, from e.g. Kepler's Laws to the Lorenz attractor.
However, finding patterns in highly multidimensional data requires automated tools.
Here, we would like to understand whether the data retrieved form the NOMAD Archive can be grouped into clusters of data that share a similar representation, where data points within the same cluster are similar to each other while being different from data points belonging to other clusters. The notion of similarity in the discussed unsupervised-learning task is strictly related to the representation of the data, here a set of atomic properties of the constituent material. 

A plethora of different clustering algorithms has been developed in the last years, each with different ideal applications (see, e.g., our tutorial notebook introducing the most popular clustering algorithms\cite{clustering-notebook}).
Among the various algorithms currently available, we chose a recent algorithm, which we will briefly outline below, that stands out for simplicity, quality of the results, and robustness.

The clustering algorithm that is employed in this notebook is the hierarchical density-based spatial clustering of applications with noise (HDBSCAN)\cite{hdbscan}, a recent extension of the popular DBSCAN algorithm\cite{dbscan}.
As density-based algorithms, HDBSCAN relies on the idea that clusters are islands of high-density points separated by a sea of low-density points.
The data points in the low-density region are labeled as `outliers' and are not associated with any clusters.
Outlier identification 
is at the core of the HDBSCAN algorithm, which uses the {\em mutual reachability distance}, i.e. a specific distance metric to distort the space so as to ``push'' outliers away from the high density regions.


Cluster definition is to some extent subtle, as many possible different combinations are acceptable.
One of the main challenges is represented by nested clusters, where it is not always trivial to decide whether a relatively large cluster should be decomposed into more subclusters, or if instead a unique supercluster should be taken.
The HDBSCAN algorithm performs a hierarchical exploration that evaluates possible subdivisions of the data into clusters.
Initially, for low values of the distance threshold, there is only one large cluster that includes all points.
As the threshold is lowered, the cluster can eventually split into smaller subclusters.
This algorithm automatically decides whether to split the supercluster, and this decision is based on how robust --- with respect to further divisions --- the new subclusters would be. 
If, for example, after a cluster division many other splittings would shortly follow while lowering the threshold distance, then the larger supercluster is taken; if, otherwise, the subclusters do not immediately face further subdivisions, they are selected instead of the large supercluster.

{\bf Dimension reduction: the {\em Visualizer}}

\begin{figure}
    \centering
    \includegraphics[width=8.5cm]{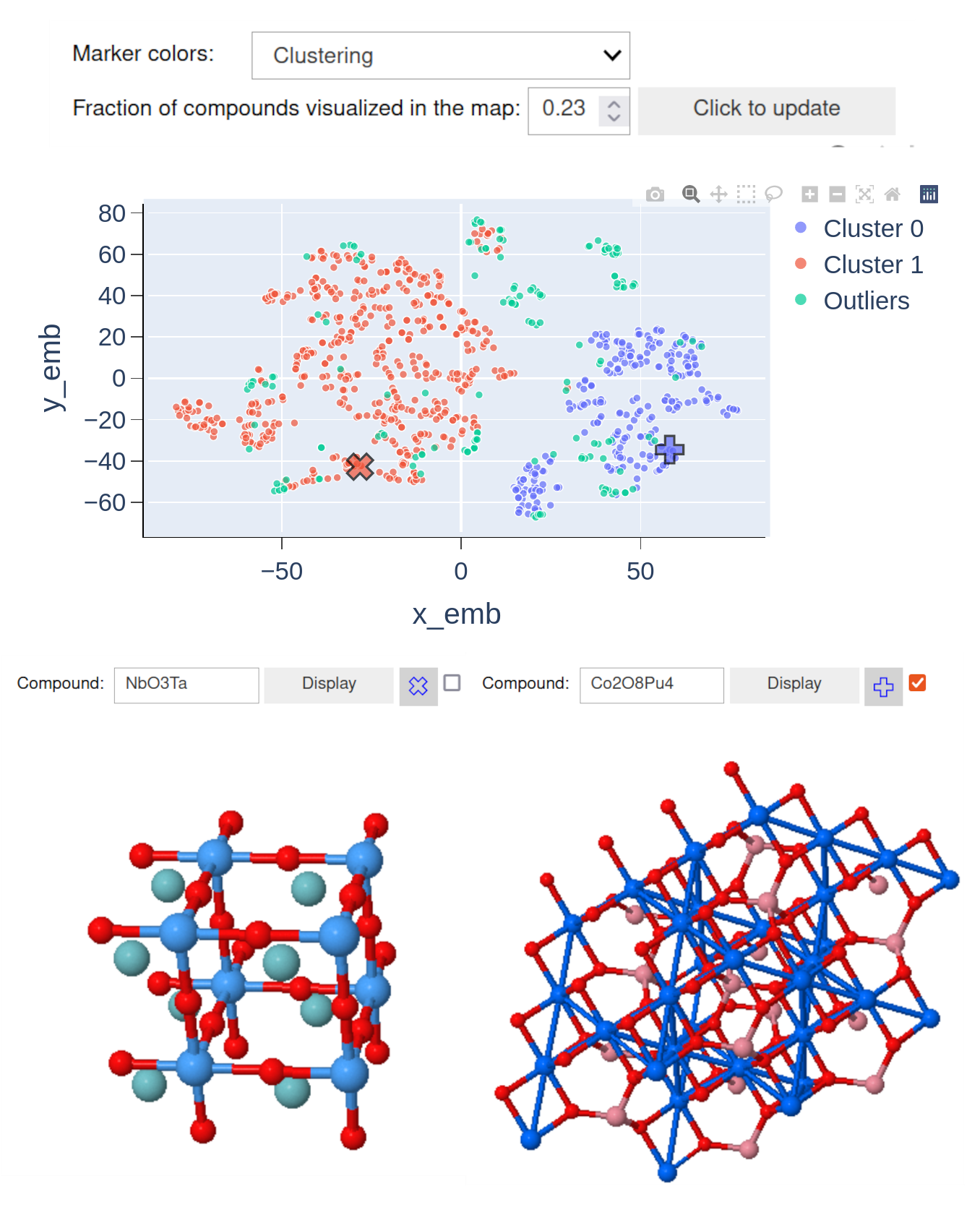}
    \caption{Snapshot of the {\em Visualizer} in the `Querying the Archive and performing Artificial Intelligence modeling' notebook. The visualization of a two-dimensional map allows to identify subsets (in AI nomenclature: clusters) of materials with similar properties. Two windows at the bottom of the map allow to view the structures of the compounds in the map. Clicking a point shows the structure of the selected material.  Ticking the box on top of the windows selects which one of the two windows is used for the next visualization. The two windows have different types of symbols (here, crosses) to mark the position on the map. It is also possible to display a specific material chosen from the Compound text box to show its structure and its position on the map, which is then labelled with a cross. In this figure, two compounds are visualized, and it is possible to spot the position of the materials on the map. }
    \label{fig:unsup_visualizer}
\end{figure}
\begin{figure}
    \centering
    \includegraphics[width=9cm]{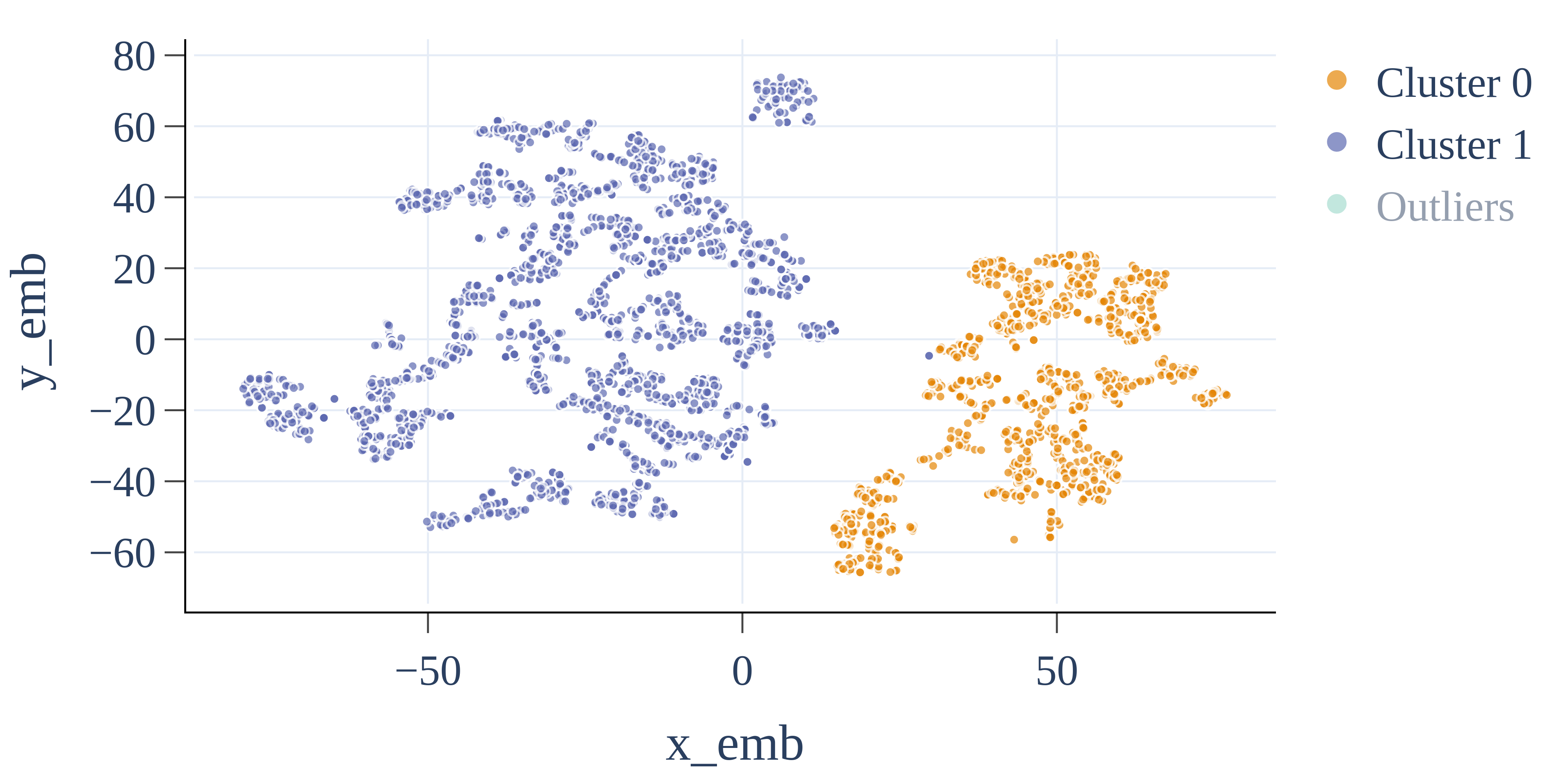}
    \caption{An example of a high-quality plot that can be produced using the visualizer. The `Toggle on/off plot appearance utils' button displays a number of controls that can be used to modify and generate the plots. It is possible to change resolution, format file, color palette for the markers, text format and size, and markers' size.}
    \label{fig:unsup_plot}
\end{figure}

The NOMAD AI Toolkit also comes with a {\em Visualizer}, a package which allows a straightforward analysis of tabulated data that contain materials structures, and which is optimized for data retrieved from the NOMAD Archive.
The visualizer is built using the \textsc{Plotly} package\cite{plotly}, which allows the creation of an interactive map, whose usability is improved using ipywidgets.
The map shows with distinct colors different clusters of materials, that were embedded into a two-dimensional plane using the dimension reduction algorithm t-SNE \cite{tSNE}. 
We would like to remark that axes in this embedding do not have a meaning, and cannot be expressed as a global function of the features spanning the original space.
This embedding algorithm, as many nonlinear embedding algorithms, finds a low dimensional representation where pairwise distances between data points are preserved, which makes it possible to visualize clusters of points in a two-dimensional plot.

Clicking on any of the points in the map displays the atomic structure of the material in one of the windows at the bottom of the map.
The position of the compound that is displayed is marked with a cross on the map.
There are two different display windows to facilitate the comparison of different structures, and the window for the next visualization is selected with a tick box on top of the visualizer.
By clicking `Display' the structure of the material and its position on the map are shown.
We also provide some plotting utilities to generate high-quality plots.
Controls for fine-tuning the printing quality and appearance are displayed by clicking the `For a high-quality print \ldots' button.

{\bf Discovering of new topological insulators: application of SISSO to alloyed tetradymites}

As a second, complementary example, we discuss a notebook that addresses an analysis of topological semiconductors\cite{cao2020artificial}.  
The employed AI method is SISSO (sure-independent screening combined with sparsifying operator \cite{ouyang2018sisso}), which combines symbolic regression with compressed sensing. In practice, for a given target property of a class of materials, SISSO identifies a low-dimensional descriptor, out of a huge number of candidates (billions, or more). The candidate descriptors, the materials genes, are constructed as algebraic expressions, by combining mathematical operators (e.g., sums, products, exponentials, powers) with basic physical quantities, called primary features. These features are properties of the materials, or their constituents (e.g., the atomic species in the material's composition), that are (much) easier to evaluate (or measure) than the target properties that are modeled by using the SISSO-selected features as input and with the mathematical relationship identified as well by SISSO. 
In Ref. \cite{cao2020artificial}, the materials' property of interest was the classification between topological vs trivial insulators.

\begin{figure}[t!]
    \centering
    \includegraphics[width=8.5cm]{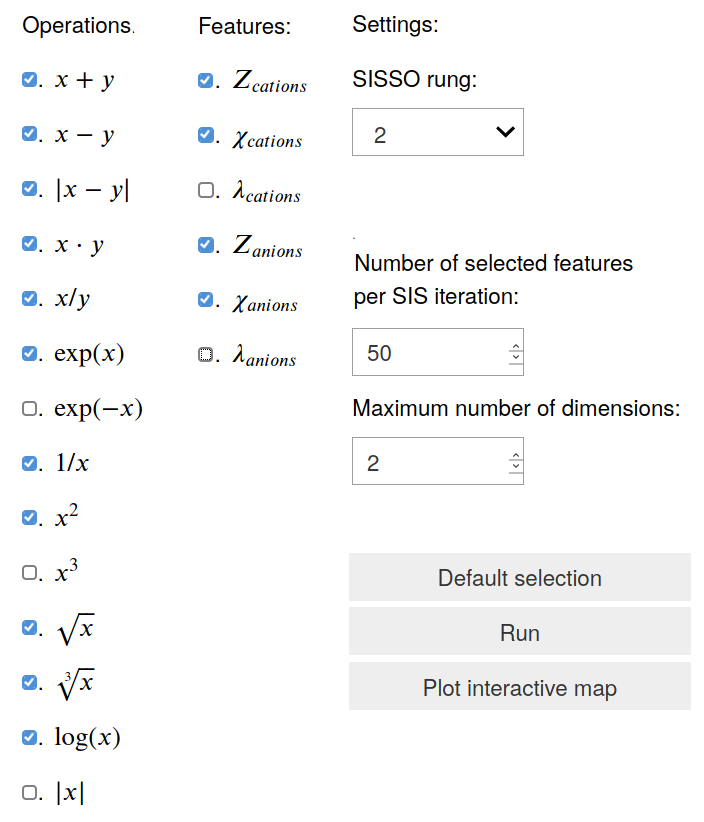}
    \caption{Graphical input interface for the SISSO training of tetradymite-materials classification, taken from the `Discovery of new topological insulators in alloyed tetradymites' notebook.}
    \label{fig:input_mask}
\end{figure}

\begin{figure}[t!]
    \centering
    \includegraphics[width=1.0\columnwidth]{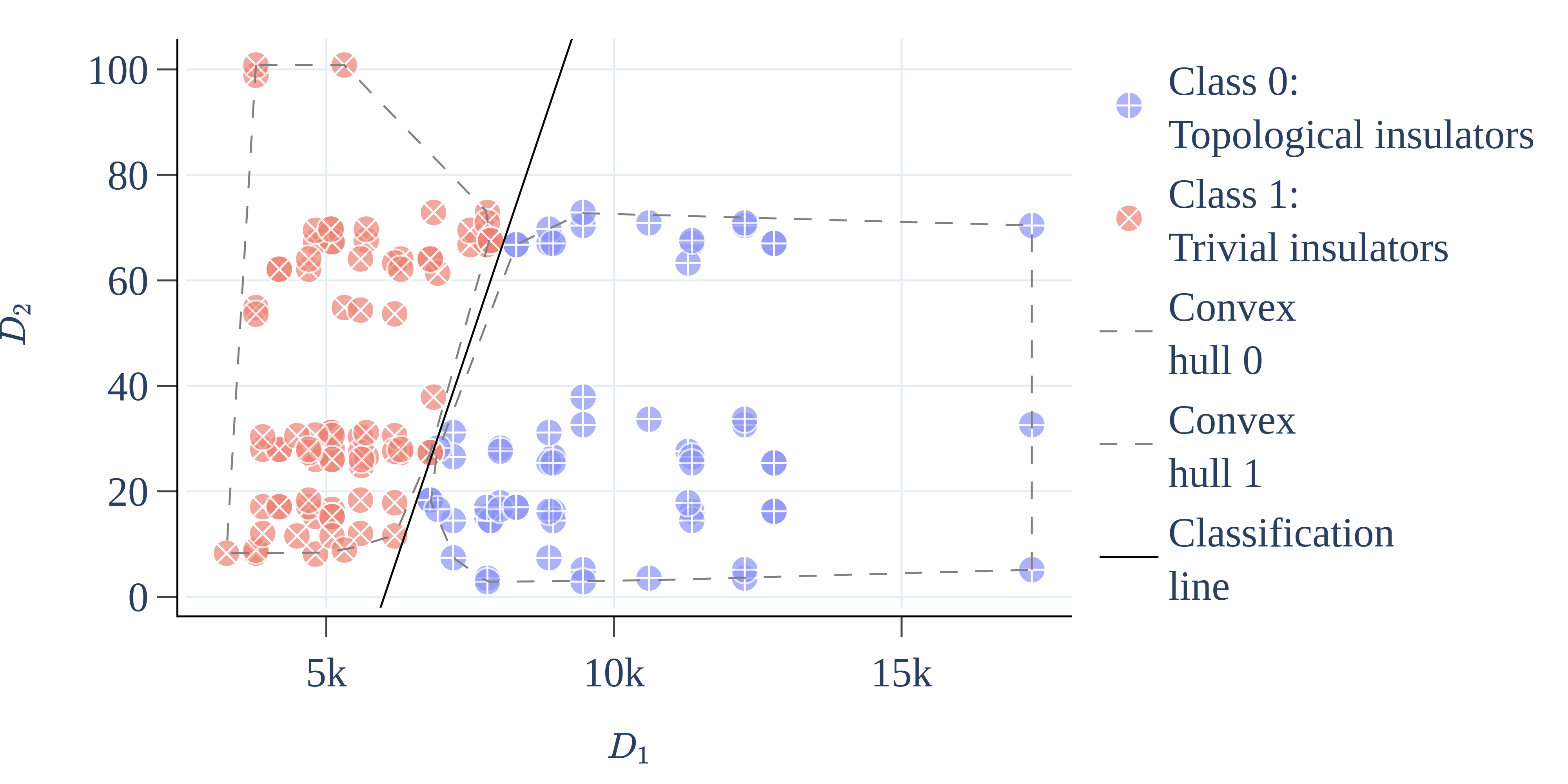}
    \caption{Interactive map of tetradymite materials, as produced with the AI-Toolkit visualizer. The topological (trivial) insulator training points are marked in red (blue). All materials falling in the convex hulls delimited by the dashed line enveloping the red (blue) points are predicted to be topological (trivial) insulators. The axes, $D_1$ and $D_2$ are the components of the descriptor identified by SISSO, in terms of analytical function of the selected input parameters (see Ref. \cite{cao2020artificial} and the notebook \cite{tetrad-notebook} for more details).}
    \label{fig:tetrad_plot}
\end{figure}

The addressed class of materials was the tetradymites family, i.e., materials with the general chemical formula $AB-LMN$, where the cations $A,B \in \{ \textrm{As, Sb, Bi} \}$ and the anions $L,M,N \in \{ \textrm{S, Se, Te} \}$, and a trigonal (R3m) symmetry. Some of these materials are known to be topological insulators and the data-driven task was to predict the classification into topological vs trivial insulators of all possible such materials, just by knowing their formula, by using as training data a set of 152 tetradymites for which the topological invariant $Z_2$ is calculated via DFT for the optimized geometries.

In the notebook `Discovery of new topological insulators in alloyed tetradymites' \cite{tetrad-notebook}
, we invite the user to interactively reproduce the results of Ref. \onlinecite{cao2020artificial}, namely the materials property map as shown in Fig. \ref{fig:tetrad_plot}. The map is obtained within the notebook, after selecting as input settings the same primary features and other SISSO parameters as used for the publication. In Fig. \ref{fig:input_mask}, we show a snapshot of the input widget, where users can select features, operators, and SISSO parameters according to their preference and test alternative results. When clicking `Run', the SISSO code is running within the container created for the user at the NOMAD server.
In the notebook, the map as shown in Fig. \ref{fig:tetrad_plot} is managed by the same {\em Visualizer} as described in Section III for the query-and-analyse notebook. This means that by mouse hovering the chemical formula of the compound represented by the marker is shown in a tooltip. By clicking a marker, the crystal structure of the corresponding material is shown in a box below the plot.

In summary, with the notebook `Discovery of new topological insulators in alloyed tetradymites', we provide an interactive, complementary support to Ref. \cite{cao2020artificial}, where the user can reproduce the results of the paper starting with the same input, by using the same code, and by going as far as re-obtaining exactly the same main result plot (except for the different graphical style). More than what can be found in the paper, the user can change the input settings to the SISSO learning, explore the results by changing the visualization settings, and browsing the structures of the single data points. The user can also use the notebook as a template and start from other data, retrieved from the NOMAD Archive, to perform an analysis with the same method, etc. 

\section{Discussion}
We presented the NOMAD AI Toolkit, a web-browser-based platform for performing AI analysis of materials-science data, both online, on NOMAD servers, and locally on own computational resources, even behind firewalls. The purpose of the AI toolkit is to provide the tools for exploiting the Findable and AI Ready (F-AIR) materials-science data that are contained in the NOMAD Repository and Archive, as well as several other databases in the field. The platform provides integrated access, via Jupyter notebooks to state-of-the-art AI methods and concepts. Shallow learning curve hands-on tutorials are provided, in the form of interactive Jupyter notebooks, for all the available tools. A particular focus is on the reproducibility of AI-based workflows associated with high-profile publications: The AI Toolkit offers a selection of notebooks demonstrating such workflows, so that users can understand step by step what was done in publications and readily modify and adapt the workflows to their own needs. We hope this example could be an inspiration to augment future publications with similar hands-on notebooks. This will allow for enhanced reproducibility of data-driven materials science papers and dampen the learning curve for newcomers to the field. 
The community is invited to contribute more notebooks in order to share cutting-edge knowledge in an efficient and scientifically robust way.

\section{Data availability}
Data used in this study are openly accessible on the NOMAD Artificial-Intelligence toolkit at \url{https://nomad-lab.eu/aitoolkit}.

\section{Code availability}
Codes used in this study are openly accessible on the NOMAD Artificial-Intelligence toolkit at \url{https://nomad-lab.eu/aitoolkit}, see in particular Refs. \cite{query-archive-notebook} and \cite{tetrad-notebook} for the codes (notebooks) of the specific examples discussed in this paper.

\section*{Acknowledgements}
We would like to acknowledge Fawzi Mohammed, Angelo Ziletti,  Markus Scheidgen, and Lauri Himanen for inspiring discussions. \\
This work received funding from the European Union’s Horizon 2020 research and innovation program under the grant agreement Nº 951786 (\mbox{NOMAD} CoE), the ERC Advanced Grant TEC1P (No. 740233), and the German Research Foundation (DFG) through the NFDI consortium `FAIRmat', project 460197019.



\bibliography{bib}

\end{document}